# Nitrogen Precooling Heat Exchanger replacement and control system upgrade in Superfluid Cryoplant at CMTF

**J. Subedi, B. Hansen, M. White, V. Patel, J. Makara, O. Atassi and G. Johnson**

Fermi National Accelerator Laboratory, Batavia, IL 60510

Email: jsubedi@fnal.gov

**Abstract**. Liquid Nitrogen precooling is used in most Cryoplants to achieve cooldown to 80 K temperature range. In one such system at Fermilab's CMTF Superfluid Cryoplant, where the Helium supply directly exchanges heat with liquid Nitrogen, freezing of Nitrogen occurred inside the heat exchanger due to heat exchanger imbalance during a Cryoplant trip. Trapped vapor pockets of N2 within the frozen heat exchanger channels were formed while warming up the heat exchanger, creating high localized pressure and subsequent damage/rupture of the heat exchanger. Replacement of the heat exchanger was done, and modifications were made in the system to rectify future occurrences. The control system was updated to bypass the heat exchanger entirely if the incoming Helium stream temperature drops below 76 K. This was done by repurposing two control valves as heat exchanger bypass valves that were previously used for a redundant 80 K adsorber in the coldbox. Additional modifications were made to further prevent return of large amount of cold Helium gas from cold end during abrupt Cryoplant shutdown. This modification has ensured high reliability of heat exchanger with prevention of freezing of Nitrogen which can damage the heat exchanger.

## 1. Introduction

The cryogenic system of Fermilab's Cryomodule Test Facility (CMTF) is supported by a state-of-the-art Superfluid Cryogenic Plant (SCP) from Linde Process Plants (LPP) which provides the refrigeration needs of this facility. Procured in 2013, this Cryoplant provides 40 K, 5 K and 2 K flows to two test areas at CMTF. It currently supports Cryomodule Test Stand 1 (CMTS1) for LCLS-II HE project which is used to test 1.3 GHz cryomodules and PIP-II injection Test Stand (PIP-II IT) program to test the front end of the future PIP-II Linac and production cryomodules [1,2]. CMTF utilizes four 60 g/s Mycom compressors to supply high pressure Helium to the SCP coldbox. 2 K operations can be achieved either by string of cold compressors that brings test stand to 2 K or a vacuum pump utilizing Tuthill Kinney warm vacuum compressor skids. This Cryoplant uses liquid Nitrogen for cooling down ambient temperature Helium to around 80 K. Addition of liquid nitrogen precooling increases the Cryoplant liquefaction capacity significantly in expense of Nitrogen consumption. The Nitrogen system is designed to directly exchange heat with incoming compressor discharge Helium to the Cryoplant. Liquid level is maintained in the liquid Nitrogen vessel. The cooldown of incoming Helium is done by latent heat of Nitrogen in E3120 heat exchanger and the boil off gas exits through another recuperative heat exchanger E3115 which cools down ambient temperature Helium as shown in *Figure 1*. Near ambient gaseous nitrogen exits the coldbox during normal operation after exchanging heat with incoming discharge Helium in this recuperative heat exchanger.

## 2. Incident

On March 30th, 2018, the Superfluid Cryoplant coldbox shut down due to low storage pressure in the middle of the evening hours. The heater for the liquid Helium dewar was set at a manual setting of 1000 W to force recovery of gaseous Helium back to storage just before the shutdown. After coldbox shutdown, various valves are programmed to automatically go to their cold box shutdown positions. Turbine inlet valves slowly ramp down to prevent sudden thrust forces that can damage the turbine. With liquid Helium dewar return valve to coldbox open and manual heat input of 1000 Watts on the dewar, a large amount of cold Helium gas continued to return up the shell side of plant's heat exchangers until the heater was turned off. This caused imbalance of flow across heat exchangers and resulting in rapid cooldown of the HP helium flow in the first heat exchanger block E3110 to below N2 freezing temperatures. As a result, the high-pressure Helium flow leaving the first heat exchanger (E3110) rapidly dropped in temperature below LN2 freezing point (< 63 K) before entering the LN2 precooling heat exchanger E3120. Nitrogen within the heat exchanger channels began to freeze. This was not immediately identified by the responding operators and Nitrogen remained below freezing temperature overnight and resulted in Heat exchanger channels being frozen solid with Nitrogen.

The next morning, in preparation for coldbox restart, the valve configuration was modified by an operator to flow warm Helium gas through the Helium side of liquid Nitrogen precooling Heat Exchanger. The frozen Nitrogen evaporated non-uniformly and gas pockets developed in the middle of the frozen channels with no path to escape. This resulted in local over pressurization causing a rupture in heat exchanger channels of E3120 causing Helium to Nitrogen leak.

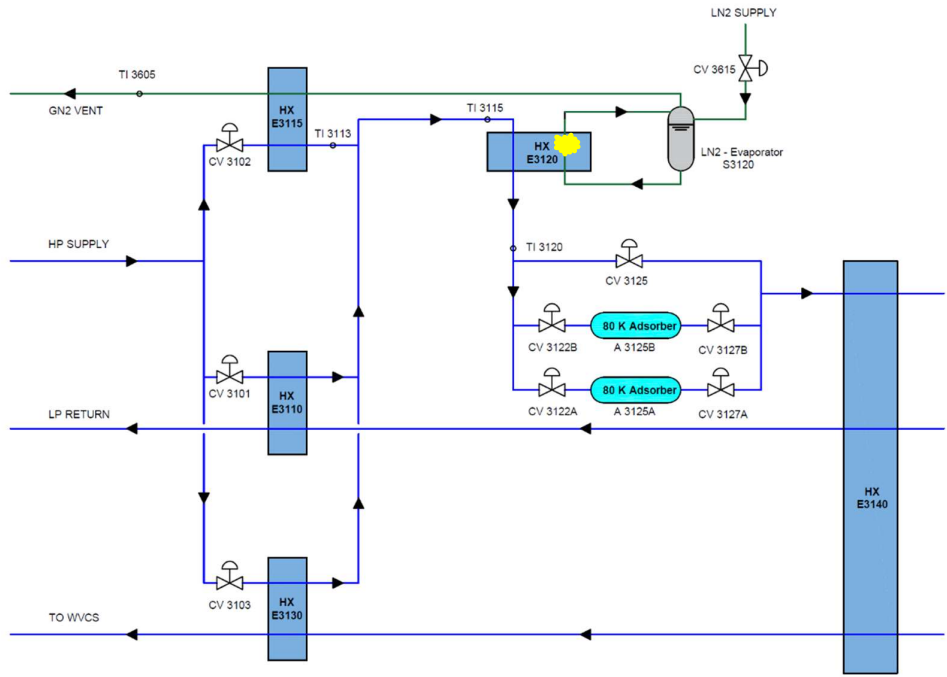

Figure 1: CMTF coldbox Nitrogen precooling simplified flow schematic before incident.

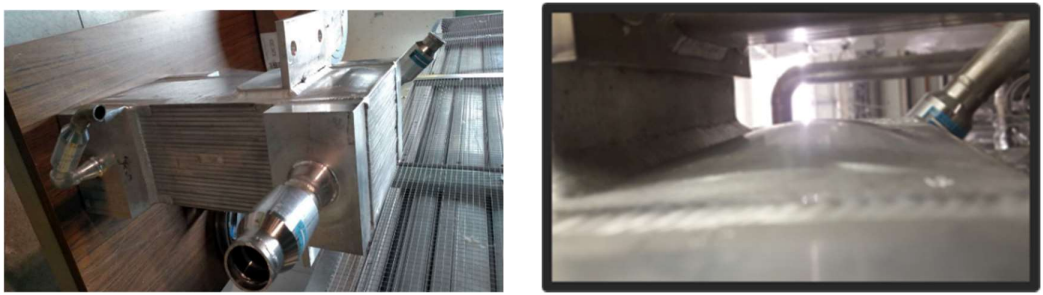

Figure 2: Images of damaged heat exchanger after the incident

### 3. Modification before transition to upgraded system

Once the leakage of the heat exchanger E3120 was confirmed, it was removed from the coldbox. The procurement of new heat exchanger had significant lead time and Cryoplant needed operation to support critical LCLS-II and PIP-II IT projects deadline. In place of the heat exchanger, a 4000 W inflow heater and two 1425 W band heaters were added on Nitrogen circuit to generate Nitrogen boil-off to provide a means of precooling Helium with the recuperative heat exchanger E3115. Over temperature protection was used for heater bands and a temperature sensor was installed in the heat exchanger Helium outlet. Table 1 below shows designed Cryoplant capacity at different configurations. While not an efficient solution, the heaters configuration provided a way to get the system operational with acceptable capacity to keep the cryogenic test programs on schedule. The addition of these heaters gave ~43% more liquefaction capacity to Cryoplant than without use of any Nitrogen precooling. The Cryoplant was operated in this configuration for 2 years until the replacement heat exchanger was delivered and a shutdown window became available to install the final replacement.

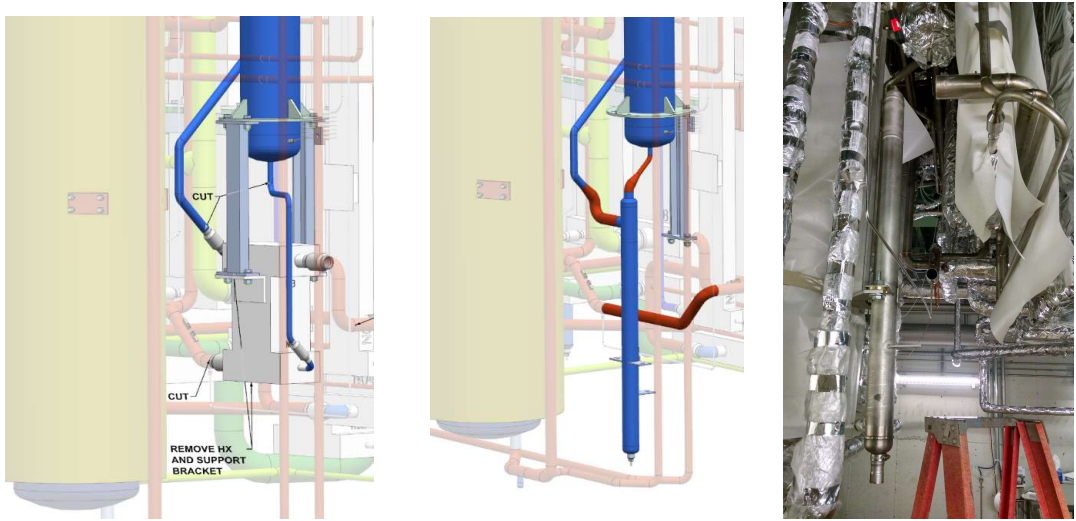

Figure 3: 3D model showing before and after incident configuration of Nitrogen precooling and actual image showing installation of pipe spool with inflow heater.

| Configuration | 40 K shield capacity | 5 K shield capacity | Liquifaction capacity |
|---|---|---|---|
| With precooling heat exchanger | 735 W | 105 W | 19.8 g/s |
| Without precooling heat exchanger | 400 W | 75 W | 7.64 g/s |
| With 7 kW heaters | 400 W | 75 W | 11 g/s |

Table 1: 40 K shield, 5 K shield and liquefaction capacity of Cryoplant at different configurations

### 4. Heat exchanger replacement and control system upgrade

A heat exchanger of similar effectiveness and capacity as the original damaged heat exchanger was procured from a vendor. This heat exchanger was installed in the coldbox as replacement for E3120 heat exchanger. Mere replacement of heat exchanger would not have prevented the catastrophic incident from reoccurring. Thus, series of pipe reconfiguration and control system upgrades were made to ensure that the incident did not repeat. Due to spatial constraint and valves requirement, one of the 80 K adsorbers in the SCP coldbox was removed and the inlet and outlet valves in the adsorber were repurposed to be used to bypass Helium flow around the heat exchanger in case of low temperature HP supply temperature. These valves CV3122A and CV3127A are controlled by single controller TC3115. The controller regulates Helium inlet temperature to the heat exchanger process variable. The controller is inverse PID controller and CV3122A, bypass valve to heat exchanger is controlled by this controller output. CV3127A, the valve which supplies Helium through the heat exchanger is set to complement CV3122A such that total valve opening is always 100 %. The control logic is as follows:
- CV3122A position = controller output
- CV3127A position = 100 % - controller output

Figure 5 below shows where the valve position should be if PID controller response is linear to the process variable.

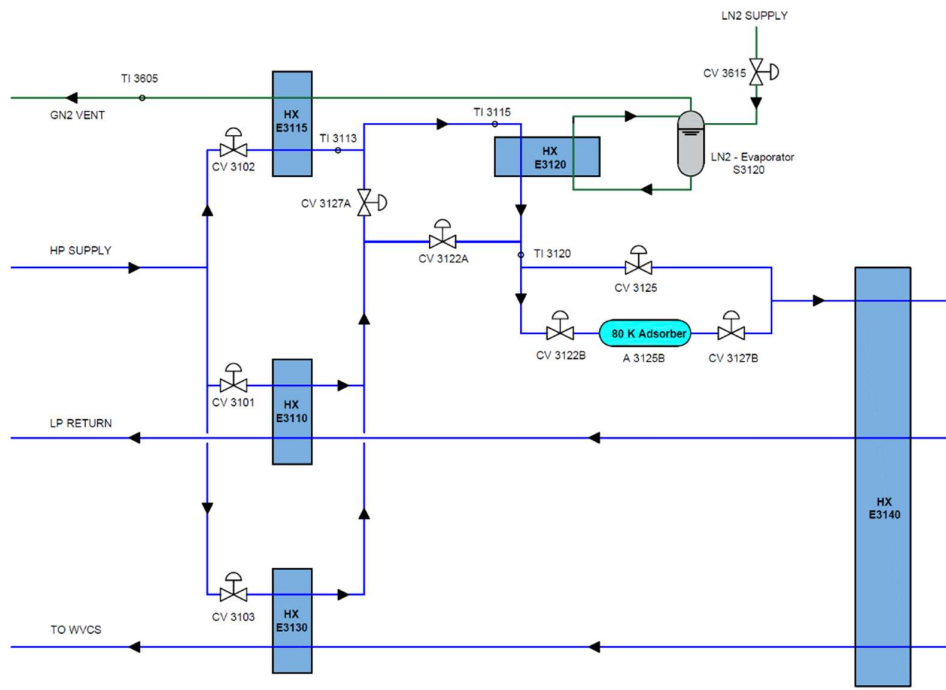

Figure 4: CMTF coldbox Nitrogen precooling simplified flow schematic after controls upgrade.

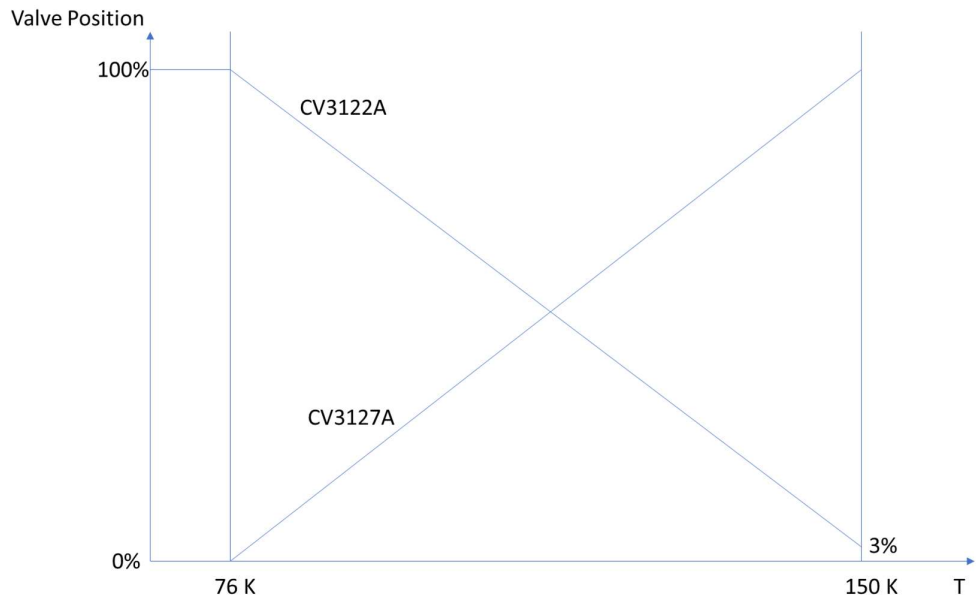

Figure 5: E3120 bypass control valves configuration with linear PID controller response.

In addition to this controller, when the Heat Exchanger Helium inlet temperature TI3115 is less than 76 K, LN2 bypass interlock is activated. CV3122A fully opens and CV3127A fully closes when this LN2 bypass mode is activated. LN2 bypass interlock resets when TI3115 and TI3120 are greater than 76 K. When LN2 bypass interlock is not activated, the controller TC3115 is enabled. This prevents heat exchange between cold Helium and Nitrogen at value close to Nitrogen freezing temperature.

The issue of cold Helium returning to the warm end was also rectified through controls upgrade. The Helium dewar return valve to coldbox was configured to be 0% after coldbox shutdown. Cooldown valve for dewar PVCD instead opens which returns boil off Helium gas directly to compressor suction, bypassing the coldbox. The dewar heater was also configured to interlock off in case of coldbox shutdown.

## 5. Conclusion

Controls upgrade after replacement of Nitrogen precooling heat exchanger has prevented the incidence of Nitrogen freezing in the heat exchanger from reoccurring. Use of two valves configuration to control and bypass flow through Nitrogen precooling heat exchanger has ensured better control of Helium inlet temperature and more stable coldbox operations. This has also increased reliability of the Cryoplant. CMTF SCP is now operational with the designed capacity and serving test requirement of LCLS II and PIP II cryomodules.

**Acknowledgments**

This manuscript has been authored by Fermi Research Alliance, LLC under Contract No. DE-AC02-07CH11359 with the U.S. Department of Energy, Office of Science, Office of High Energy Physics. The authors wish to recognize the dedication and skills of APS-TD/Cryogenics technical staff involved in the installation and commissioning of this system.